\documentclass[prd, preprint, nofootinbib]{revtex4}
\usepackage{graphicx}
\usepackage{amsfonts}
\usepackage{latexsym}
\usepackage{amssymb}

\begin{document}

\title{Dilution of axion dark radiation by thermal inflation}

\author{Hironori Hattori, Tatsuo Kobayashi, Naoya Omoto}
 \affiliation{Department of Physics, Hokkaido University, Sapporo 060-0810, Japan}

\author{Osamu Seto}
 \affiliation{Department of Life Science and Technology,
  Hokkai-Gakuen University, Sapporo 062-8605, Japan}

%
\begin{abstract}
Axions in the Peccei-Quinn (PQ) mechanism provide a promising solution
 to the strong CP problem in the standard model of particle physics. 
Coherently generated PQ scalar fields could dominate the energy density
 in the early Universe and decay into relativistic axions,
 which would conflict with the current dark radiation constraints.
We study the possibility that a thermal inflation driven
 by a $U(1)$ gauged Higgs field dilutes such axions.
A well-motivated extra gauged $U(1)$ would be the local $B-L$ symmetry. 
We also discuss the implication for the case of $U(1)_{B-L}$ and
 an available baryogenesis mechanism in such cosmology.
\end{abstract}

\pacs{}
\preprint{HGU-CAP-037} 
\preprint{EPHOU-15-0009}

\vspace*{3cm}
\maketitle


\section{Introduction}

Inflationary cosmology has been the paradigm in modern cosmology~\cite{Yamaguchi:2011kg}.
The accelerating expansion in the early Universe not only solves
 the flatness and horizon problem~\cite{Inflation}
 but also provides the density perturbations which grow to the large-scale
 structure~\cite{InflationFluctuation}.
The false vacuum energy inducing quasi-de Sitter expansion during inflation
 is transferred into
 the energy of radiation with high temperature through reheating processes, so that
 the filling of early Universe with high-density and high-temperature thermal plasma
 is realized.

It is often expected that such a hot universe directly connects to
 the hot big bang universe
 in which big bang nucleosynthesis (BBN) takes place after one second.
However, there are various possibilities of nontrivial evolution scenarios
 where a stage dominated by temporal matter or vacuum energy would exist 
 between the primordial inflation and the onset of BBN.
In addition, through the evolution of the early Universe from
 a very high-energy to low-energy region,
 it is supposed that the Universe has undergone a series of cosmological phase transitions
 and the symmetry breaking of particle theory, e.g.,
 a large-gauge (sub)group of grand unified theory (GUT), Peccei-Quinn (PQ) symmetry~\cite{Peccei:1977hh},
 and the electroweak gauge symmetry in the standard model (SM) of particle physics.
There are two well-motivated symmetries expected to be broken at an intermediate scale 
 from a particle physics viewpoint.
One is a PQ symmetry introduced to solve the strong CP problem in the SM.
The other is $B-L$ symmetry which might be gauged symmetry at a high-energy scale
 and its breakdown would generate Majorana masses of right-handed (RH) neutrinos
 to account for neutrino oscillation phenomena~\cite{Mohapatra:1980qe}.

A scalar field to break the PQ symmetry has small couplings to SM particles suppressed
 by the PQ breaking scale $f_a$, also known as the axion decay constant.
The PQ scalar field with a ``wine bottle'' potential 
 could be produced as coherent oscillation due to its scalar nature
 and temporally dominate the energy density of the Universe
 if its decay rate is very small because of suppressed couplings.
The radial direction of the PQ scalar~\footnote{From now on, we simply call it the PQ scalar.}
 would mostly decay into axions or SM particles through loop processes.
For the KSVZ axion model~\cite{KSVZ} where heavy quarks carrying a PQ charge are
 indeed heavy enough, this could happen.
For the DFSZ axion model~\cite{DFSZ} where
 axion multiplets have direct couplings with the SM Higgs field,
 this could be the case if its coupling between the PQ field and the SM Higgs field is small enough
 so that the PQ radial scalar decay into Higgs bosons is suppressed.
Thus, associated with the breakdown of the global $U(1)_{PQ}$ symmetry,
 the PQ scalar likely decays into mostly axions in this model and
 cannot reheat the Universe well.
Those overproduced massless axions act as dark radiation which is
 nowadays stringently constrained.
The Planck data show the effective number of neutrino spices~\cite{Planck:2015xua}
\begin{equation}
 N_{eff} = 3.15 \pm 0.23 ,
\end{equation}
 and the difference from the SM prediction is the so-called $\Delta N_{eff}$.

This dark radiation axion is an unwanted relic and needs to be diluted anyway.
Thermal inflation is a well-known mechanism to dilute such unwanted relics~\cite{Lyth:1995ka}
 and is driven by a scalar field $\varphi$, often called the flaton,
 whose scalar potential $V(\varphi)$ is of the symmetry-breaking type with a large VEV.
In a hot Universe after the reheating of the primordial inflation,
 the scalar field $\varphi$ is thermally excited and stays at the origin $\varphi=0$
 because the thermal correction in the effective potential makes
 the origin of the potential
 to a temporal minimum with its large false vacuum energy $V(0)$.
When the false vacuum energy becomes comparable to the background radiation
 energy density, the Universe starts accelerating expansion again.
Since the radiation energy density is redshifted away as its temperature
 decreases exponentially,
 the origin $\varphi=0$ becomes unstable and the inflation terminates.
With the enormous entropy production by the $\varphi$ decay,
 unwanted relics can be diluted away.

In this paper, we investigate the condition of successful thermal inflation
 to dilute axions generated by the late decay of the dominated PQ scalar field. 
If this flaton $\varphi$ is a gauge singlet and has an (approximate) global $U(1)$,
 then the axions associated with the flaton could be produced again as shown in Ref.~\cite{AK}.
Thus, in order not to have an axion overproduction problem after thermal inflation,
 we assume that a flaton field is charged under a local $U(1)$ symmetry.
We also discuss the implication in the case that this local $U(1)$ symmetry 
 is identified with gauged $U(1)_{B-L}$.

\section{Thermal inflation in an axion-dominated Universe}

\subsection{Thermal inflation by a wine bottle potential}

We describe the outline of the scenario with a thermal inflation period
 driven by canonical scalar field $\varphi$,
 which is responsible for breaking its local $U(1)$ symmetry.
For illustrative purposes, in this section,
 we assume the scalar potential for thermal inflation
 is of the wine bottle form.
The scalar potentials for $\varphi$ are given by
\begin{eqnarray}
 V(\varphi) &=& \lambda_{\varphi}(v^2-|\varphi|^2)^2 .
\label{Pot:wb}
\end{eqnarray}
A flaton field $\varphi$ is assumed to
 be in thermal equilibrium through interactions with particles in the hot thermal bath
 and, hence, the thermal correction; in fact, the thermal mass term,
\begin{equation}
 \delta V = \frac{g_{\varphi}}{24} T^2 |\varphi|^2,
\label{thermalmass}
\end{equation}
 with $T$ being the temperature of the thermal plasma, is added in the scalar potential.  
Here, $g_{\varphi}$ is parametrizing the coefficient, while
 sometimes we may use an effective coupling with another particle $h \equiv \sqrt{g_{\varphi}}$
 instead of $g_{\varphi}$ in the rest of this paper.

Before evaluating the number of $e$-folds in a relativistic axion-dominated Universe,
 we here review how to estimate it in the standard radiation-dominated Universe.
When the false vacuum energy $V(0)$ dominates over the energy density of
 the background radiation $\rho = \pi^2 g_* T^4/30 $ with $g_*$ being
 the number of relativistic degrees of freedom, a thermal inflation starts. 
We define the initial temperature of a thermal inflation by 
\begin{equation}
 \frac{\pi^2 g_*}{30} T_i^4 = V(0).
\label{Cond:Inf:ini}
\end{equation}
A thermal inflation terminates when the temperature drops below the critical temperature $T_C$
 defined by 
\begin{equation}
 \frac{g_{\varphi}}{24} T_C^2 = \lambda_{\varphi} v^2,
\end{equation}
 and the origin $\varphi=0$ becomes unstable. In the rest of this paper,
 we use the final temperature $T_f$ instead of $T_C$.
We find the initial and final temperature, and the number of $e$-folds as
\begin{eqnarray}
T_i &=& \left(\frac{\pi^2}{30}g_*\right)^{-1/4}\left(\lambda_\varphi v^4 \right)^{1/4}, \label{Ti} \\
T_f &=& 4\sqrt{\frac{3\lambda_\varphi}{g_\varphi}} v , \label{Tf} \\
N_4 &=& -\ln{4\sqrt{3}}-\frac{1}{4}\ln\left(\frac{\pi^2}{30} g_* \right)
 -\frac{1}{4}\ln\frac{\lambda_\varphi}{g_\varphi^2} \label{N}.
\end{eqnarray}
Here, the subscript $4$ in $N_4$ indicates, for later convenience, 
 that the scalar potential is quartic.
We find several number of $e$-folds by thermal inflation can be realized only for 
\begin{equation}
 \lambda_\varphi \ll g_\varphi^2 , \label{Cond:lam-g}
\end{equation}
 from Eq.~(\ref{N}).
One may find that $g_\varphi$ would be expressed as
\begin{equation}
 g_\varphi \sim \lambda_\varphi + g^2+ \sum y^2 
\end{equation}
where $g$ and $y$ stand for the gauge coupling of $\varphi$ for gauged $U(1)$
 and Yukawa couplings, respectively, to a fermion $\psi$ as in 
\begin{equation}
 {\cal L} \subset y \bar{\psi} \varphi \psi .
\end{equation}

On the other hand, the scalar self-coupling $\lambda_\varphi$ also receives
 corrections like  
\begin{equation}
 \lambda_\varphi \rightarrow
 \lambda_\varphi + \frac{1}{64 \pi^2}\sum( \lambda_\varphi^2 + g^4- \sum y^4 ),
 \label{LoopLambda}
\end{equation}
 with $\sum$ denoting the summation with respect to degrees of freedom.
Therefore, in fact, the condition (\ref{Cond:lam-g}) can be satisfied
 only if accurate cancellations happen in Eq.~(\ref{LoopLambda}).
Realization of a thermal inflation by the potential (\ref{Pot:wb})
 is unstable against radiative correction,
 and in this sense, successful thermal inflation by this potential is difficult.
This would be a reason why Lyth and Stewart have considered
 the flaton potential lifted by higher-order nonrenormalizable terms
 in the original ``thermal inflation'' paper~\cite{Lyth:1995ka}, as we will also do.

Before turning to higher-order potential,
 we estimate the number of $e$-folds in the axion-dominated Universe
 by the dominated PQ scalar field decay, because the above
 Eqs.~(\ref{Ti}), (\ref{Tf}), and (\ref{N}) have been derived
 in a radiation-dominated Universe background.
In the KSVZ model, the PQ scalar mostly decays mostly into
 two axions and a little into SM particles.
We use $Br$ to parametrize the decay branching ratio of the PQ scalar
 into SM particles. 
Then we have
\begin{eqnarray}
&& \rho_{total} = \rho_{SM\,rad}+\rho_{axion} , \\
&& \rho_{SM\,rad} : \rho_{axion} = Br : 1-Br , \\ 
&& \rho_{SM\,rad} = \frac{\pi^2}{30}g_* T^4 . 
\end{eqnarray}
Since thermal inflation starts in the axion-dominated Universe,
 the condition $\rho_{total} \simeq\rho_{axion} \simeq V(\varphi=0)$ is rewritten as
\begin{equation}
 \frac{1}{Br}\frac{\pi^2 g_*}{30} T_i^4 = V(0).
\label{Cond:Inf:ini2}
\end{equation}
We obtain 
\begin{eqnarray}
T_i &=& \left(\frac{1}{Br} \frac{\pi^2}{30}g_*\right)^{-1/4}
 \left(\lambda_\varphi v^4 \right)^{1/4}, \label{Tia}\\
N_4 &=& -\ln{4\sqrt{3}}-\frac{1}{4}\ln\left(\frac{\pi^2}{30} g_* \right)
 -\frac{1}{4}\ln\left(\frac{1}{Br} \right)
 -\frac{1}{4}\ln\frac{\lambda_\varphi}{g_\varphi^2} .\label{Na}
\end{eqnarray}
In the axion-dominated Universe,
 the temperature interval between $T_i$ and $T_f$ becomes shorter
 because the energy density of thermalized radiation is subdominant.
As a result, the number of $e$-folds also becomes small, which can been seen
 as the effect of the third term in Eq.~(\ref{Na}).
We list various physical quantities about thermal inflation in Table~\ref{Table:wb}.
One can easily find the hierarchy of inequality~(\ref{Cond:lam-g})
 for realizing even just $N \sim 1$.

%
\begin{table}[!ht]
\caption{Quantities in thermal inflation by the potential (\ref{Pot:wb})}
\begin{tabular}{|c|c|c|c|c|c|c|c|}\hline
$\lambda$&$h$&$v$(GeV) &$T_i$(GeV) &$T_f$(GeV) &$N$&$\Delta N_{eff}$&$T_R$(GeV) \\ \hline
$10^{-6}$&$2.54$&$10^{8}$&$7.43\times10^5$&$2.73\times10^5$&$1.00$&$0.05$&$1.6\times10^{6}$\\ \hline
\end{tabular}
\label{Table:wb}
\end{table}

\subsection{Thermal inflation by a higher-power potential}

As we have seen,
 thermal inflation by a wine bottle potential can realize the expansion
 of only ${\cal O}(1)$ $e$-folds if the condition (\ref{Cond:lam-g}) is satisfied.
Hence, in this subsection, as a reference, 
 we estimate the number of $e$-folds in the case where the potential
 is lifted by a nonrenormalizable higher-order term.
The scalar potential is expressed as
\begin{equation}
V(\varphi) = V_0 - m^2|\varphi|^2 +\frac{|\varphi|^{2n}}{\Lambda^{2(n-2)}}.
\label{Pot:higher}
\end{equation}
Here, we stress that the absence of the $|\varphi|^4$ term is not necessary.
We just assume that it is negligible due to the small coupling constant.
$V_0$, the VEV, and the mass of $\varphi$ are expressed as
\begin{eqnarray}
 V_0 &=& (n-1)\frac{v^{2n}}{\Lambda^{2(n-2)}} , \\
 v &=& \langle\varphi\rangle = \left(\frac{m^2 \Lambda^{2(n-2)}}{n}\right)^{\frac{1}{2(n-1)}} , \\
 m_{\varphi}^2 &=& n(n-1)\frac{v^{2(n-1)}}{\Lambda^{2(n-2)}} ,
\end{eqnarray}
by using the stationary condition and $V(v)=0$. 
The number of $e$-folds by thermal inflation is give by
\begin{eqnarray}
N_{2n}
&=&N_4-\frac{1}{4}\ln\frac{n^2}{4(n-1)}+\frac{1}{2}(n-2)\ln\left(\frac{M_P}{v}\right) , \\
N_4 &=&-\ln4\sqrt{3}-\frac{1}{4}\ln\left(\frac{\pi^2}{30}g_*\right)+\frac{1}{2}\ln\frac{\Lambda}{M_P}h ,
\end{eqnarray}
 with $M_P$ being the reduced Planck mass.
We list various physical quantities about thermal inflation in Table~\ref{Table:higher}.
Thermal inflation can be realized with more sensible coupling $h$ than
 that in Table~\ref{Table:wb}. 

%
\begin{table}[!ht]
\caption{Quantities in thermal inflation by the potential (\ref{Pot:higher})}
\begin{tabular}{|c|c|c|c|c|c|c|c|}\hline
$\Lambda$(GeV) &$h$&$v$(GeV) &$T_i$(GeV) &$T_f$(GeV) &$N$&$\Delta N_{eff}$&$T_R$(GeV) \\ \hline
$10^{16}$&$8.27\times10^{-3}$&$10^{8}$&$2.79\times10^3$
&$1.03\times10^3$&$1.00$&$0.05$&$5.9\times10^{3}$\\ \hline
$10^{16}$&$8.27\times10^{-2}$&$10^{10}$&$2.79\times10^6$
&$1.03\times10^6$&$1.00$&$0.05$&$5.9\times10^{6}$\\ \hline
$10^{16}$&$8.27\times10^{-1}$&$10^{12}$&$2.87\times10^9$&$1.03\times10^9$&$1.00$&$0.05$&$5.9\times10^{9}$\\ \hline
\end{tabular}
\label{Table:higher}
\end{table}

\section{Relic abundances}

\subsection{Axion dark radiation}

As we have seen, if the PQ scalar field dominates the energy density of the Universe, 
 its decay produces many axions, and the Universe ends up
 with relativistic axion domination.
When the total energy density $\rho_{total}$ from dominated axion $\rho_a$ and
 subdominant radiation $\rho_{rad}$
 becomes comparable with $V(\varphi)$, $t=t_i$,
 the thermal inflation begins.
Then we have
\begin{eqnarray}
V(\varphi=0) = \rho_{total}(t_i) = \frac{1}{Br}\rho_{rad}(t_i) 
 = \frac{1}{Br}\rho_{rad}(t_f)\left(\frac{a(t_f)}{a(t_i)}\right)^4 .
\label{V2rho_a}
\end{eqnarray}
Here $t_f$ stands for the time when the thermal inflation ends.
After the thermal inflation, 
 $\varphi$ decays into SM particles and potentially non-SM particles again. 
Let us define the decay branching ratio into non-SM particles $br$ for later convenience.
By using Eq.~(\ref{V2rho_a}), the ratio of axion to radiation is expressed as
\begin{equation}
\rho_{rad}|_{H=\Gamma} = (1-br) e^{4N} \rho_a|_{H=\Gamma}\left(\frac{a|_{H=\Gamma}}{a|_{H(t_f)}}\right) ,
\end{equation}
 where $N$ and $\Gamma$ are the number of $e$-folds of the thermal inflation
 and the decay rate of $\varphi$.
The resultant axion dark radiation contribution is expressed in terms of $\Delta N_{eff}$ as
\begin{equation}
\Delta N_{eff} = \frac{43}{7}\left(\frac{43/4}{g_*}\right)^{1/3} \times 
 \left.  \frac{\rho_a}{\rho_{rad}}\right|_{H=\Gamma} .
\end{equation}

\subsection{Reheating temperature and possible baryogenesis scenarios}

The reheating temperature after inflation is estimated as
\begin{equation}
T_R = \left( \frac{90}{\pi^2 g_*(T_R)} (1-br) \Gamma^2 M_P^2 \right)^{1/4} .
\label{Formula:TR}
\end{equation}
In the following,
 we estimate a reference value of the reheating temperature 
 under the assumption of the instantaneous reheating $\Gamma = H(t_f)$, which gives the highest reheating temperature.
The availability of baryogenesis mechanisms depends on the reheating temperature after thermal inflation $T_R$.

For high enough reheating temperature $T_R \gtrsim 10^9$ GeV,
 thermal leptogenesis by the lightest heavy RH neutrino decay of those with hierarchical masses is
 one of the simplest scenarios of baryogenesis~\cite{FY,Buchmulleretal}.

Nonthermal leptogenesis by RH neutrinos with hierarchical masses is available for
 a reheating temperature $10^9$ GeV $ \gtrsim T_R \gtrsim 10^6$ GeV~\cite{NonTherLept}.
If this local $U(1)$ is in fact the gauged $U(1)_{B-L}$ symmetry, $\varphi$ is identified
 with the Higgs field to break this symmetry with the $B-L$ charge $2$, and
 the decay $\varphi$ into two RH neutrinos $N_R$ is nothing but nonthermal production of $N_R$.

For $T_R \lesssim 10^6$ GeV,
 low-scale thermal leptogenesis requires an enhancement of CP violation.
Here, for information, we refer to two examples.
One possibility is the so-called resonant leptogenesis, where
 two RH neutrino masses are strongly degenerated and CP violation is enlarged
 due to RH neutrino self-energy~\cite{Pilaftsis:2003gt}.
Another way is an extension of the Higgs sector. It has been shown that, 
 in the so-called neutrinophilic Higgs model,
 large enough CP violation can be obtained for
 the lightest RH neutrino mass of ${\cal O}(10^4)$ GeV~\cite{Haba:2011ra}.

Another promising scenario would be electroweak baryogenesis.
For a recent review, see, e.g., Ref.~\cite{Morrissey:2012db}.

\subsection{Results}

\begin{figure}[htbp]
 \begin{minipage}{0.49\hsize} 
  \begin{center}
\includegraphics[width=80mm]{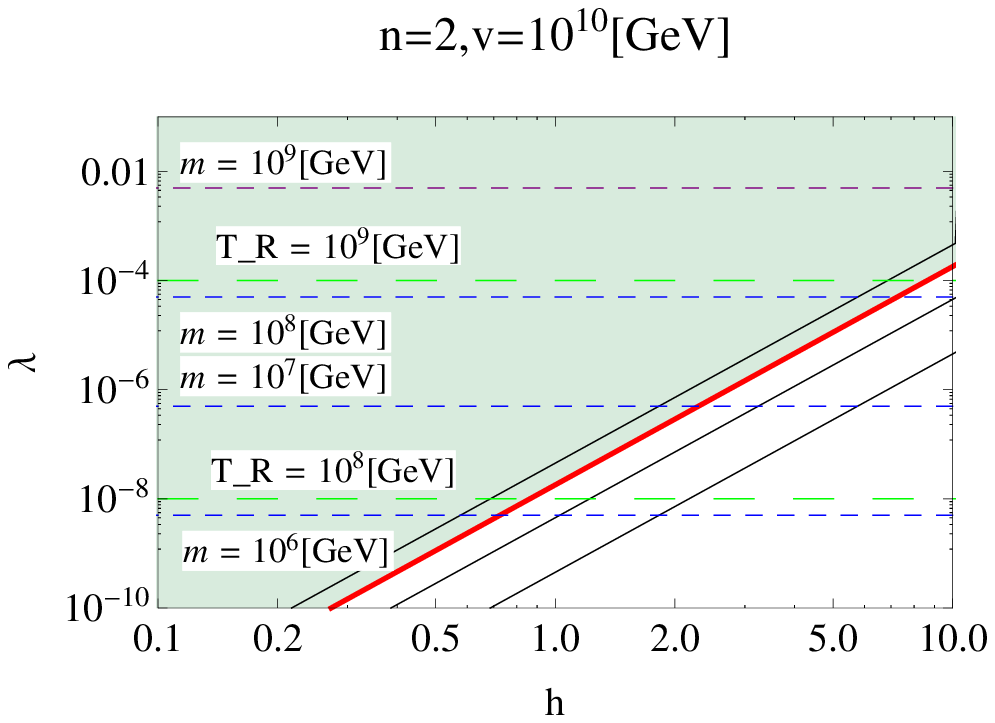}
  \end{center}
\end{minipage}
\hfill
 \begin{minipage}{0.49\hsize} 
  \begin{center}
\includegraphics[width=80mm]{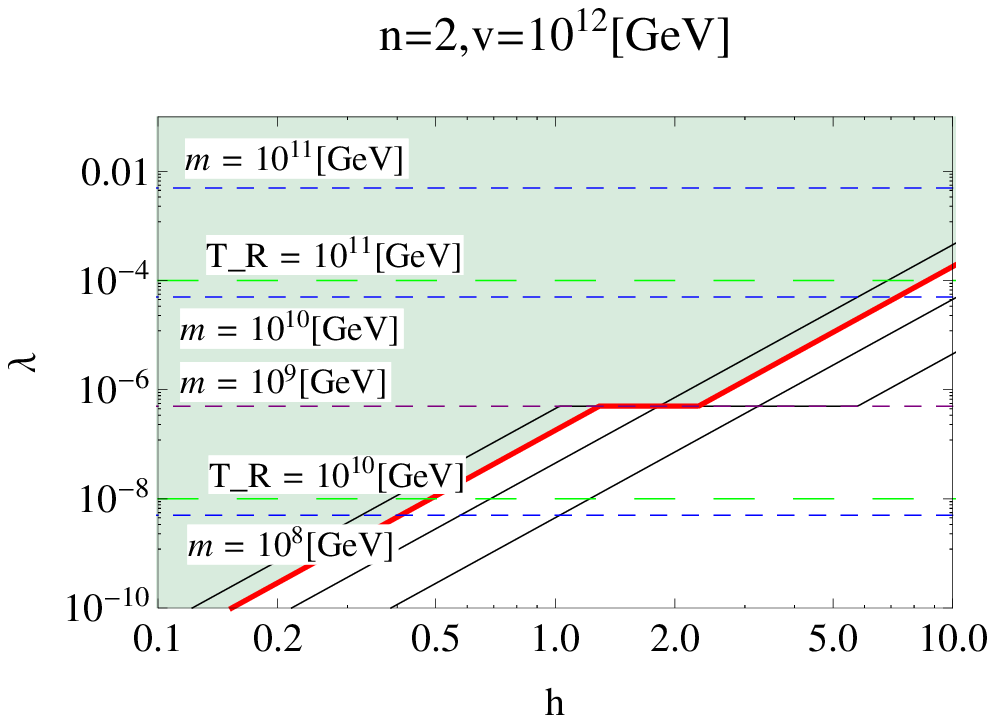}
  \end{center}
\end{minipage}
\caption{
Contours of the resultant $\Delta N_{\rm eff} = 1, 0.4$(thick red)$, 0.1$, and $0.01$ with solid lines from left to right, the mass of $\varphi$ with dashed lines and
 the possible maximal reheating temperature after thermal inflation $T_R$ with long dashed lines
 by the potential (\ref{Pot:wb}). The shaded region corresponds to $\Delta N_{eff} > 0.4$ which is
 disfavored by the Planck (2015) data.}
\label{Fig:deltaN4}
\end{figure}

\begin{figure}[htbp]
 \begin{minipage}{0.49\hsize} 
  \begin{center}
\includegraphics[width=80mm]{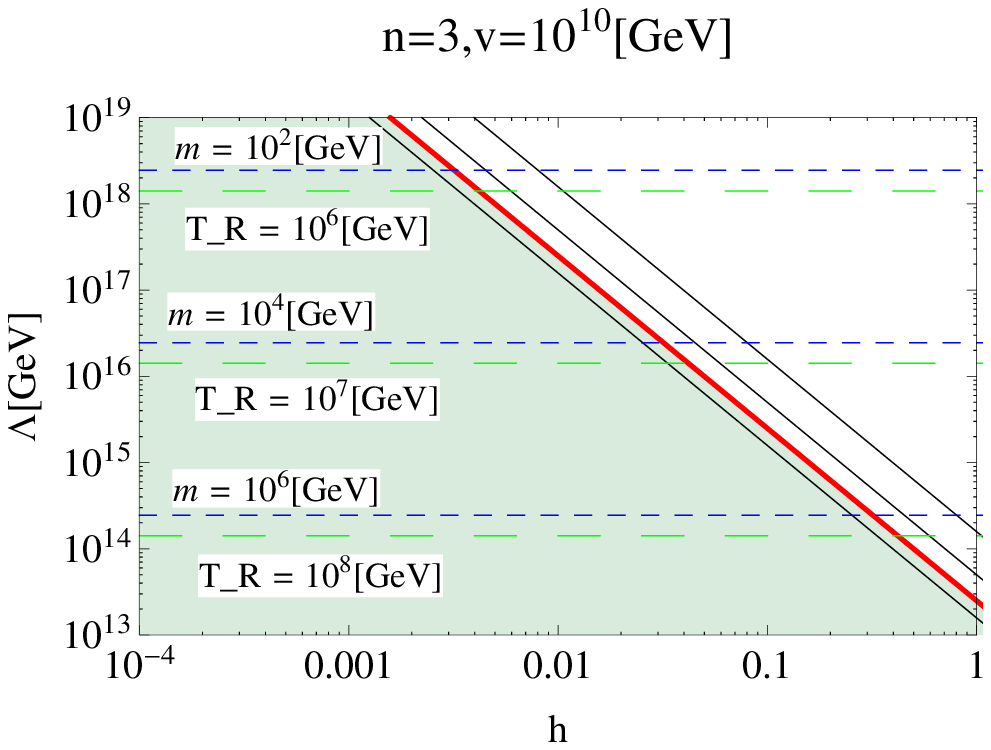}
  \end{center}
\end{minipage}
\hfill
 \begin{minipage}{0.49\hsize} 
  \begin{center}
\includegraphics[width=80mm]{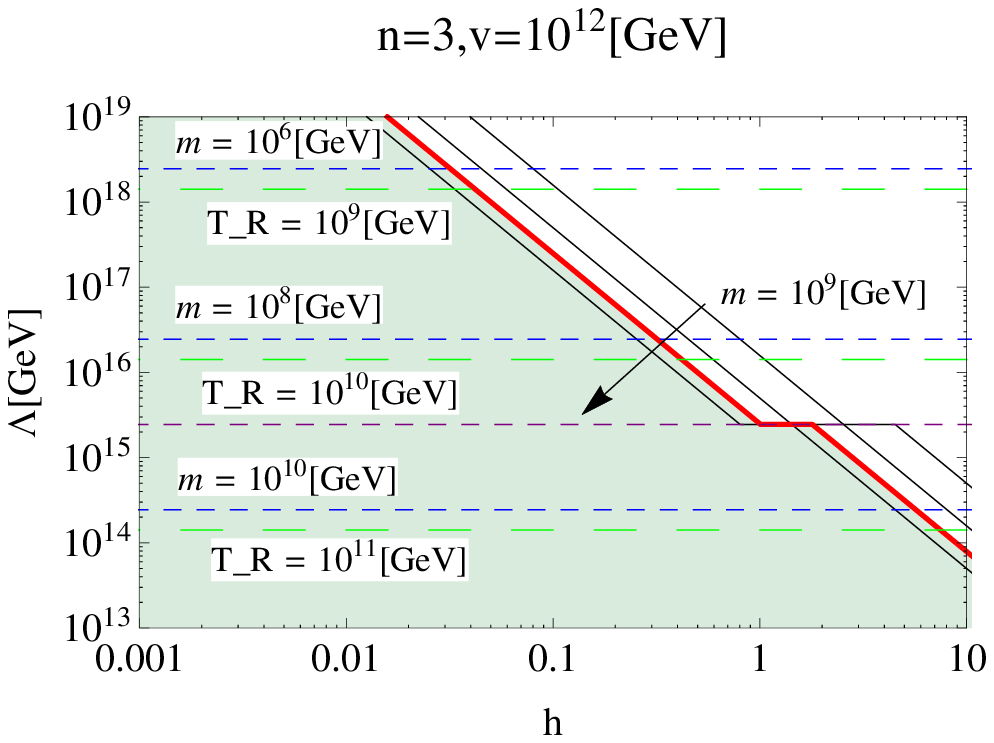}
  \end{center}
\end{minipage}
\caption{The same as Fig.~\ref{Fig:deltaN4} but
 for the potential (\ref{Pot:higher}) with $n=3$. }
\label{Fig:deltaN6}
\end{figure}

Here, by using Figs.~\ref{Fig:deltaN4} and \ref{Fig:deltaN6},
 we summarize the viable parameter space and available baryogenesis mechanism
 for some benchmark points we have studied.
Inflation with the potential (\ref{Pot:wb}) is labeled as $n=2$, and $n=3$ indicates inflation
 by the potential (\ref{Pot:higher}) with $n=3$.
Although we comment on cases with $n=2$,
 one should remember that realization of thermal inflation by the $n=2$ potential
 is, in general, difficult as stated above, so this is only for information purpose. 
In order to have a large enough CP violation $\varepsilon \gtrsim 10^{-6}$ in the $N_R$ decay,
 which is equivalent to the so-called Davidson-Ibarra bound of the lightest RH neutrino mass
 for leptogenesis $M_{N_R} > 10^9$ GeV~\cite{LowerBound,Davidson:2002qv},
 we suppose $M_{N_R} \simeq 10^9$ GeV and that the decay $\varphi \rightarrow N_R N_R$ is kinematically forbidden
 for $m_{\varphi} < 10^9$ GeV.
We consider two cases of the PQ scalar VEV, $v= 10^{10}$ and $10^{12}$ GeV.
As far as implications for baryogenesis are concerned, 
 the conclusion is the same for $v\lesssim 10^{10}$ GeV.

\subsubsection{$n=2, v=10^{12}$ GeV case}

For most of the parameter space, we have $T_R > 10^9$ GeV.
Thus, the standard thermal leptogenesis could work.

\subsubsection{$n=2, v=10^{10}$ GeV case}

For most of the parameter space, $10^9$ GeV $> T_R > 10^6$ GeV is realized;
 however, $m_{\varphi} \lesssim 10^9$ GeV.
Nonthermal leptogenesis by the $\varphi$ decay 
 does not work as long as a hierarchical mass is assumed.
Thus, a low-scale thermal leptogenesis with an enhanced CP violation
 or the electroweak baryogenesis with the extension of the Higgs sector is needed.

\subsubsection{$n=3, v=10^{12}$ GeV case}

For most of the parameter space, we have $T_R > 10^9$ GeV.
Thus, the standard thermal leptogenesis could work.

\subsubsection{$n=3, v=10^{10}$ GeV case}

As for the $n=2, v=10^{10}$ GeV case, $T_R > 10^6$ GeV is realized;
 however, $m_{\varphi} \lesssim 10^9$ GeV.
Nonthermal leptogenesis by the $\varphi$ decay 
 does not work because the $\varphi$ decay into them is kinematically forbidden.
Thus, a low-scale thermal leptogenesis with an enhanced CP violation
 or the electroweak baryogenesis with the extension of the Higgs sector is needed.


\section{Summary}

In this paper we have investigated scenarios with successful thermal inflation
 by a gauged $U(1)$ Higgs flaton field
 to dilute axions generated by late decay of the dominated PQ scalar field which 
 dominantly decays into relativistic axions. 
We find the available parameter space to dilute axions.
By estimating the flaton Higgs boson mass and
 the possible highest reheating temperature after thermal inflation, 
 if the $U(1)$ symmetry is the gauged $U(1)_{B-L}$,
 we find that a promising viable baryogenesis in this cosmology
 is not nonthermal leptogenesis with hierarchical RH neutrino masses
 but high- or low-scale thermal leptogenesis or electroweak baryogenesis.


\section*{Acknowledgments}
This work was supported in part by the Grant-in-Aid for Scientific Research 
No.~25400252 (T.K.) and on Innovative Areas No.~26105514 (O.S.)
 from the Ministry of Education, Culture, Sports, Science and Technology in Japan. 
%


\appendix



\end{document}